\documentstyle[preprint,revtex,eqsecnum]{aps}
\begin{document}
\draft
\begin{flushright}
SFU-HEP-102-93 \\
\end{flushright}
\vskip.3in
\begin{title}
 {\bf Self-mass for Massive Quark}
\end{title}
\author{Zheng Huang and K.S.\ Viswanathan}
\begin{instit}
Department of Physics, Simon Fraser University,
 Burnaby, B.C., Canada V5A 1S6
\end{instit}
\begin{abstract}
We study the behavior of the self-mass for a quark with a
 current mass
larger than $\Lambda_{\mbox{\scriptsize QCD}}$, as a function of its  Euclidean
momentum and mass, in QCD. An expression for the Bethe-Salpeter
kernel of
the Schwinger-Dyson (SD) equation valid in both the infrared
 and ultraviolet regions is obtained based on a
renormalization group analysis. The resulting SD equation is
solved numerically. It is found that the quark constituent mass
at zero momentum is substantially enhanced
 due to its effective gauge interaction. The solution in the
 ultraviolet region agrees well with the known asymptotic solution.
 The self-mass scales exactly as the on-shell current mass at
 a fixed momentum.
\end{abstract}

\narrowtext

\section{Introduction}
Dynamical chiral symmetry breaking in QCD has been studied
extensively \cite{1} and yet has not been completely
understood. A rigorous proof of this phenomenon based on a
reliable non-perturbative scheme has not so far been given. The
standard method to study this problem is to start with the
Schwinger-Dyson (SD) equation \cite{2}. However, some
difficulties arise when applying this equation. The most
serious one is the use of a single gluon exchange (the ladder
approximation). Although the SD equation incorporates non-perturbative
features,
the ladder approximation of the integral kernel is essentially a
perturbative scheme and its validity must be justified by the
smallness of the effective gauge coupling.

Some efforts have been made in order to establish the validity
of the ladder approximation by making use of the asymptotic
freedom in QCD. Instead of the complete solution of the SD
equation, the asymptotic solutions when $p^2\rightarrow
\infty$ have been discussed where the effective coupling is
small \cite{3}. Two types of solutions, the irregular and
 regular ones, have been found by linearizing the differential
SD equation in the ultraviolet region. In the presence of a
current mass, a linear combination of both solutions satisfies
the UV boundary condition while the irregular solution is
dominant in the UV region. It is then speculated that the
regular solution which goes to zero faster than the irregular
one in the UV region
may become important, and perhaps becomes dominant in the infrared
region if the current mass is small. As a result, the light
quark may acquire a dynamical mass much larger than its current
 mass in the IR region. This has been referred to as the
dynamical chiral symmetry breaking. However, the above statement
stays as a speculation as long as we do not have an appropriate
approximation to make the SD equation solvable in the IR
region. Some attempts on this issue have been made by assuming a certain
behavior of the effective coupling constant in the IR region
\cite{4}. In general, there has not been a complete solution to
the SD equation valid both in the UV and in the IR regions with
approximations based on QCD.

In this paper,  we study a slightly different aspect of
the solution to the SD equation however approaching to the same problem:
the self-mass for a very massive quark. When the quark current
mass is much larger than $\Lambda_{\mbox{\scriptsize QCD}}$, a renormalization
group equation
(RGE) allows us to derive an approximation to the integral
kernel valid both in the UV and in the IR regions. A complete solution
to the non-linear SD equation is then obtained by  numerical
means. To observe how the gauge interaction affects the result,
we rescale all dimensionful quantities such as  the self-mass,
the momentum and the current mass by the current mass. As an
effect of the renormalization, the interaction strength and the
rescaled quark self-mass are functions of the
current mass. We find that it does go up in the IR region when
 the current mass becomes small, especially when $p^2\rightarrow
 0$. It is then expected based on extrapolation that the
 constituent mass for a light quark (defined as the self-mass at
 $p^2\rightarrow 0$) can be very large compared with its current mass.
Certainly, we
cannot quite approach the point where the quark current mass
is as small as a few MeV's. Our method is only applicable when
the current mass is bigger than $\Lambda_{\mbox{\scriptsize QCD}}$. Thus at
best the strange
quark may be accounted for in our method. We do not claim that
our result is a proof of the dynamical chiral symmetry breaking.
However, it may approach to the same limit from a different
point of view, as much as the approach based on the momentum
extrapolation from the UV region to the IR region, if not
better.  In addition, it gives a complete description of the
self-mass for the very massive quarks such as the charm and the
bottom quarks, and a picture of the transition going from the
heavy quarks to the less massive quarks.

The plan of this paper is as follows. In section \ref{theSD} we
derive the renormalized SD equation in the framework of QCD.
Section \ref{kernel} illustrates the RGE-improved integral
kernel for the massive quark. The numerical solutions for
different current masses are found and discussed in section
\ref{numer}.

\section{The Renormalized SD Equation}\label{theSD}
The standard form for the quark propagator is defined in the
momentum space as
\begin{eqnarray} \label{1}
S(p)=\frac{1}{\not\! pA(p^{2})-B(p^{2})}
\end{eqnarray}
where $A(p^2)$ is the wave function and $B(p^2)$ is referred to
as the quark self-mass. In the absence of gauge interactions
$B(p^2 )$ is equal to the quark current mass $m_0$. When the
gauge interaction is turned on, $B(p^2)$ receives corrections. In
general, $B(p^2)$ is a function of the momentum and the quark
current mass $m_0$. The quark propagator satisfies the
following Schwinger-Dyson equation
\begin{eqnarray} \label{2}
S(p)=\not\! p-m_{0}+\mbox{i}g^{2}C_{2}(N)\int
\frac{\mbox{d}^{4}k}{(2\pi )^{4}}\Gamma
^{\mu}(p,k)S(k)D_{\mu\nu}(p-k)\gamma^{\nu}
\end{eqnarray}
where $g$ is the gauge coupling constant,
$C_{2}(N)=\frac{N^{2}-1}{2N}$ for $SU(N)$ gauge group,
$\Gamma^{\mu}$ is the complete quark-antiquark vertex and
$D_{\mu\nu}$ stands for the complete gluon propagator. As we
can see from (\ref{1}), if we are to solve $S(p)$ from
(\ref{2}), $\not\! pA(p^{2})$ may dominate $B(p^{2})$ in
the UV region and we have to look for the subleading behavior
of $S(p)$ in order to determine the self-mass. This
complication may be simplified by use of the Ward identities
(for simplicity, we consider all quarks having the same bare
mass)
\begin{eqnarray} \label{3}
(p-k)^{\mu}\Gamma^{i}_{5\mu}(p,k)=-2\mbox{i}m_{0}\Gamma^{i}_{5}(p,k)
+ \gamma_{5}\frac{\lambda_{i}}{2}S^{-1}(p) +
S^{-1}(k)\frac{\lambda_{i}}{2}\gamma_{5}
\end{eqnarray}
where $\Gamma^{i}_{5\mu}$'s are the vertices of the colorless
axial-vector currents $J^{i}_{\mu 5}=\bar \psi\gamma
_{\mu}\gamma _{5}\frac{\lambda_{i}}{2}\psi$, $\Gamma^{i}_{5}$'s
are the vertices of the colorless pseudoscalar densities $J^{i}_{5}
=\bar \psi\gamma _{5}\frac{\lambda_{i}}{2}\psi$,
$\lambda_{i}$'s are the $SU(N_{f})$ matrices. The vertices
 $\Gamma^{i}_{5\mu}$ and $\Gamma^{i}_{5}$ satisfy equations of
 the Bethe-Salpeter type:
\begin{eqnarray} \label{4}
 \Gamma^{i}_{5\mu}(p,k)_{\alpha\beta}  =
 \frac{\lambda_{i}}{2}\left( \gamma_{\mu}\gamma_{5} \right)_{\alpha\beta} &+&
\int
 \frac{\mbox{d}^{4}q}{(2\pi
 )^{4}}K(p,k,q)_{\alpha\beta\alpha^{\prime}\beta^{\prime}} \nonumber \\
   &   & \left[

S(q)\Gamma^{i}_{5\mu}(q,q+p-k)S(q+p-k)\right]_{\alpha^{\prime}\beta^{\prime}}
 \; ;
\end{eqnarray}
\begin{eqnarray} \label{5}
\Gamma^{i}_{5}(p,k)_{\alpha\beta}  =
 \frac{\lambda_{i}}{2}\left( \mbox{i}\gamma_{5} \right)_{\alpha\beta} &+& \int
 \frac{\mbox{d}^{4}q}{(2\pi
)^{4}}K(p,k,q)_{\alpha\beta\alpha^{\prime}\beta^{\prime}}
  \nonumber \\
   &   & \left[
   S(q)\Gamma^{i}_{5}(q,q+p-k)S(q+p-k)\right]_{\alpha^{\prime}\beta^{\prime}}
\end{eqnarray}
where $K(p,k,q)_{\alpha\beta\alpha^{\prime}\beta^{\prime}}$ is the 2PI
fermion-antifermion scattering kernel.

Substituting (\ref{4}) and (\ref{5}) into (\ref{3}) we get
\begin{eqnarray} \label{6}
\mbox{i}(\not\! p-\not\! k)\gamma_{5} &+& \int
\frac{\mbox{d}^{4}q}{(2\pi )^{4}}K(p,k,q)\left[
S(q+p-k)\gamma_{5}+\gamma_{5}S(q)\right] \nonumber \\
&=& 2m_{0}\gamma_{5}+\gamma_{5}S^{-1}(p)+S^{-1}(k)\gamma_{5} \; .
\end{eqnarray}
By taking the limit $p-k\rightarrow 0$ and defining
$q=\frac{1}{2}(p+k)$ we obtain
\begin{eqnarray} \label{7}
\gamma_{5}B(q^{2})=m_{0}\gamma_{5}+\int \frac{\mbox{d}^{4}k}{(2\pi
)^{4}}K(q,k)
\left[ S(k)\gamma_{5}B(k^{2})S(k)\right] \; .
\end{eqnarray}
Eq.\ (\ref{7}) may be visualized by a skeleton diagram shown in
Figure \ref{fig1}.

Eq.\ (\ref{7}) stands for the most general relation which the
bare quark self-mass must satisfy. So far we have not made any
approximations. To actually solve the equation, we need to know
$A(q^{2})$ and $K(q,k)$, which, in turn, satisfy another set of
equations involving $B(q^{2})$ and $S(q)$. This seems a
hopeless circle unless we make some approximations. If the
coupling constant in $K(q,k)$ is small, we can expand the
kernel perturbatively in the sense of the Hartree-Fock
approximation. The whole point of studying the SD equation in
this perturbative scheme  is that  it still represents a
resummation of infinitely many ladder diagrams which cannot be
done in a pure perturbative calculation.

It is then clear that we have to renormalize the SD equation
such that the quantities appearing in Eq.\ (\ref{7}) become
renormalized in order to carry out the perturbative expansion.
Let us consider the renormalized functions
\begin{eqnarray}
S_{R}(k)& =&Z^{-1}_{\psi}(\mu ,\Lambda )S(k,\Lambda ) \label{8}\\
K_{R}(q,k) &=& Z^{2}_{\psi}(\mu ,\Lambda )K(q,k,\Lambda )\; ,
\label{9}
\end{eqnarray}
where the bare quantities depend on an ultraviolet cutoff
$\Lambda$, $Z_{\psi}(\mu ,\Lambda )$ is the renormalization
constant for the fermion propagator and $\mu$ is the
renormalization point. The bare current mass $m_{0}$, of
course, is also dependent on $\Lambda$. Substituting (\ref{8})
and (\ref{9}) in (\ref{7}) one then obtains the renormalized SD
equation
\begin{eqnarray} \label{10}
\gamma_{5}B_{R}(p^{2})=Z_{\psi}(\mu ,\Lambda )m_{0}\gamma_{5}+\int
\frac{\mbox{d}^{4}k}{(2\pi )^{4}}K_{R}(p,k)
\left[ S_{R}(k)\gamma_{5}B_{R}(k^{2})S_{R}(k)\right] \; .
\end{eqnarray}
where the bare mass $m_{0}(\Lambda )$ is related to the
renormalized mass in the leading-logarithmic approximation
\begin{eqnarray} \label{11}
m_{0}(\Lambda )=m(\mu )\left[ \frac{\ln (\mu^{2}/ \Lambda_{\mbox{\scriptsize
QCD}}^{2})}{\ln
(\Lambda ^{2}/ \Lambda_{\mbox{\scriptsize QCD}}^{2})}\right]^{d}
\end{eqnarray}
and $d=3C_{2}(N)/\beta_{0}$ where $\beta_{0}=11-2N_{f}/3$. It is
noteworthy that the renormalized quantities in (\ref{10}) are
in general functions of the momentum, the renormalized coupling
constant $g(\mu )$, the renormalized current mass $m(\mu )$,
the renormalized gauge parameter $\xi (\mu )$ and the
renormalization point $\mu$. $B_{R}(p^{2})$, for example, is
just a short hand notation. In below, we further drop the
subscript ``R'' but it should be understood implicitly
throughout the rest of the paper.

\section{A Bethe-Salpeter Kernel for Massive Quark}\label{kernel}
Our goal is to solve the renormalized SD equation for the
self-mass $B(p^{2})$ of the heavy quark whose current mass is
larger than $\Lambda_{\mbox{\scriptsize QCD}}$. We  expand the Bethe-Salpeter
kernel
$K(p,k)$ perturbatively if the relevant coupling is small. A
lowest few diagrams in the straightforward expansion in terms of the
renormalized coupling $g(\mu )$ are depicted in Figure \ref{fig2}
and the contributions are
\begin{eqnarray} \label{12}
K\left( p,k;g(\mu ),m(\mu ),\xi (\mu );\mu
\right)_{\alpha\beta\alpha^{\prime}\beta^{\prime}}
=
C_{2}(N)(\gamma^{\mu})_{\beta\beta^{\prime}}(\gamma^{\nu})_{\alpha\alpha^{\prime}}d_{\mu\nu}
(p-k)\nonumber\\
g^{2}(\mu )\left[ 1+{\cal O}\left( g^{2}(\mu )\ln
\frac{4\pi
\mu^{2}}{a_{1}p^{2}+a_{2}k^{2}+a_{3}(p-k)^{2}+a_{4}m^{2}}\right)\right]
\end{eqnarray}
where $d_{\mu\nu}$ is the free gluon  propagator and $a_{i}$'s
are some kinetic constants. In general, the renormalized gauge
coupling $g(\mu )$ is not small in strong interactions, the
series  $[\;\ldots\; ]$ in (\ref{12}) does not converge and
the perturbative expansion makes no sense. However, thanks to
the asymptotic freedom in QCD, $g(\mu )$ can be very small if
the renormalization point $\mu$ is chosen to be much larger
than $\Lambda_{\mbox{\scriptsize QCD}}$. Recall that in the leading-logarithmic
approximation
\begin{eqnarray} \label{13}
\frac{g^{2}(\mu )}{4\pi}=\frac{4\pi}{\beta_{0}\ln
\mu^{2}/\Lambda_{\mbox{\scriptsize QCD}}^{2}}\; .
\end{eqnarray}
The kernel at different renormalization points are related
through the RGE. Thus one can expand $K(p,k)$ at a very high
scale where the effective coupling is very small and calculate
$K(p,k)$ at a desired scale from the RGE. In the meantime, we
also have to get rid of a large multiplicative logarithmic factor
in (\ref{12}) to guarantee that the higher order terms are
negligible compared to $1$. It is thus clear that $\mu^{2}$ must
be scaled to the biggest among $p^{2}$, $k^{2}$,$(p-k)^{2}$ and
$m^{2}$ so that the logarithmic factor is small, in addition,
it must be much larger than $\Lambda_{\mbox{\scriptsize QCD}}^{2}$ so that
$g^{2}(\mu )$ is
also very small. Since there is always a factor $1/(p-k)^{2}$
in the gluon free propagator, the integral in (\ref{12}) gets
its main contribution around $(p-k)^{2}\sim 0$. Therefore we
should expand the kernel at a scale characteristic of $\max
\{ p^{2},k^{2},m^{2}\}$ which itself must be larger than
$\Lambda_{\mbox{\scriptsize QCD}}^{2}$ and use the RGE to obtain a kernel at
the physical
scale.

The RGE analysis proceeds as follows \cite{6}. The renormalized kernel
satisfies a RGE
\begin{eqnarray} \label{14}
\left\{ \mu\frac{\partial}{\partial\mu}+\beta
(g)\frac{\partial}{\partial g}+\gamma_{m}m\frac{\partial}{\partial
m}
+ \delta (g)\xi
\frac{\partial}{\partial\xi}-2\gamma_{F}(g)\right\}\nonumber\\
 K\left( p,k;g(\mu
),m(\mu ),\xi (\mu );\mu \right)=0
\end{eqnarray}
which basically tells us how the kernel changes as the
renormalization point $\mu$ changes. The solution to (\ref{14})
is known and is given by
\begin{eqnarray} \label{15}
& &K\left( p,k;g(\mu
),m(\mu ),\xi (\mu );\mu \right)\nonumber\\
& &=K\left( p,k;\bar g(t
),\bar m(t),\bar \xi (t );\mu \mbox{e}^{t} \right)\exp \left[
-2\int ^{\bar g(t)}_{g(\mu )}\mbox{d}x\frac{\gamma_{F}(x)}{\beta
(x)}\right]
\end{eqnarray}
where
\begin{eqnarray} \label{16}
t=\int ^{\bar g(t)}_{g(\mu )}\frac{\mbox{d}x}{\beta (x)}\;
&,&\quad\quad \bar m(t)=m(\mu )\left[
\int ^{\bar g(t)}_{g(\mu )}\mbox{d}x\frac{\gamma_{m}(x)}{\beta
(x)}\right]\; , \\
\bar \xi (t)&=&\xi (\mu )\left[
\int ^{\bar g(t)}_{g(\mu )}\mbox{d}x\frac{\delta (x)}{\beta
(x)}\right]\; .\nonumber
\end{eqnarray}
Let us emphasize that $\bar g(t)$, $\bar m(t)$ and $\bar \xi
(t)$ are functions of the new scale $\mu \mbox{e}^{t}$ and
$\Lambda_{\mbox{\scriptsize QCD}}$, and are {\em not} functions of $\mu$. The
other useful
relation is based on dimensional grounds when rescaling the
dimensionful parameters. For example, the following relation
holds ($t=\ln p/p_{0}$)
\begin{eqnarray} \label{17}
K\left( p,k;g(\mu
),m(\mu ),\xi (\mu );\mu \right)=\mbox{e}^{2t}K\left(
p_{0},\mbox{e}^{-t}k;g(\mu
),\mbox{e}^{-t}m(\mu ),\xi (\mu );\mbox{e}^{-t}\mu \right)\; .
\end{eqnarray}

To obtain a RGE-improved integral kernel, we choose a
rescaling factor $\mbox{e}^{t}$ where
\begin{eqnarray} \label{18}
t=\frac{1}{2}\ln \frac{\max\left\{
p^{2},k^{2},m^{2}\right\}}{\mu ^{2}}
\end{eqnarray}
and the on-shell current mass is defined by
\begin{eqnarray} \label{19}
m=m(\mu =m)\; .
\end{eqnarray}
Combining (\ref{15}) and (\ref{17}), one gets \cite{7}
\begin{eqnarray}
& &K\left( p,k;g(\mu
),m(\mu ),\xi (\mu );\mu \right)\nonumber\\
& &=K\left( p,k;\bar g(t
),\bar m(t),\bar \xi (t );\mu \mbox{e}^{t} \right)\exp \left[
-2\int ^{\bar g(t)}_{g(\mu )}\mbox{d}x\frac{\gamma_{F}(x)}{\beta
(x)}\right]\label{20}\\
& & = \mbox{e}^{2t}K\left(
\mbox{e}^{-t}p,\mbox{e}^{-t}k;\bar g(t
),\mbox{e}^{-t}\bar m(t),\bar \xi (t );\mu \right)\exp \left[
-2\int ^{\bar g(t)}_{g(\mu )}\mbox{d}x\frac{\gamma_{F}(x)}{\beta
(x)}\right]\; .\label{21}
\end{eqnarray}
Now we can calculate the effective kernel $K\left(
\mbox{e}^{-t}p,\mbox{e}^{-t}k;\bar g(t
),\mbox{e}^{-t}\bar m(t),\bar \xi (t );\mu \right)$ instead and substitute
it back in (\ref{21}). According to (\ref{18}), the largest
among $\mbox{e}^{-t}p$, $\mbox{e}^{-t}k$ and $\mbox{e}^{-t}\bar
m(t)$ is $\mu$, thus there is no large logarithmic factor in the
expansion. The effective coupling $\bar g(t)$ is given in the
leading-logarithmic approximation
\begin{eqnarray} \label{22}
\frac{\bar g^{2}(t)}{4\pi}=\frac{4\pi}{\beta_{0}\ln
\max\left( p^{2},k^{2},m^{2}\right)/ \Lambda_{\mbox{\scriptsize QCD}}^{2}}
\end{eqnarray}
which can be very small if $\max\left(
p^{2},k^{2},m^{2}\right)\gg \Lambda_{\mbox{\scriptsize QCD}}^{2}$. The
requirement that $\max\left(
p^{2},k^{2},m^{2}\right)\gg \Lambda_{\mbox{\scriptsize QCD}}^{2}$ can be
achieved if
$p^{2}\gg \Lambda_{\mbox{\scriptsize QCD}}^{2}$ or $m^{2}\gg\l^{2}$. $k^{2}$,
however, is the
integral variable which must run from $0$ to $\infty$.
Therefore, we can neglect the high order terms and write the
effective kernel to a good approximation
\begin{eqnarray} \label{23}
K\left(
\mbox{e}^{-t}p,\mbox{e}^{-t}k;\bar g(t
),\mbox{e}^{-t}\bar m(t),\bar \xi (t );\mu
\right)_{\alpha\beta\alpha^{\prime}\beta^{\prime}}& =&
C_{2}(N)(\gamma^{\mu})_{\beta\beta^{\prime}}(\gamma^{\nu})_{\alpha\alpha^{\prime}}\nonumber
\\
& & d_{\mu\nu}
(\mbox{e}^{-t}p-\mbox{e}^{-t}k)
\bar g^{2}(t)\; .
\end{eqnarray}
Substituting (\ref{23}) into (\ref{21}) we obtain a
RGE-improved kernel
\begin{eqnarray} \label{24}
K(p,k)=C_{2}(N)(\gamma^{\mu})_{\beta\beta^{\prime}}(\gamma^{\nu})_{\alpha\alpha^{\prime}}d_{\mu\nu}
(p-k)\bar g^{2}(t)\exp \left[
-2\int ^{\bar g(t)}_{g(\mu )}\mbox{d}x\frac{\gamma_{F}(x)}{\beta
(x)}\right]\; .
\end{eqnarray}

In a region where $p^{2}>m^{2}\gg\l^{2}$,
\begin{eqnarray} \label{25}
\bar g^{2}(t)=\bar g^{2}(p^{2})\theta (p^{2}-k^{2}) +
\bar g^{2}(k^{2})\theta (k^{2}-p^{2})
\end{eqnarray}
which has been discussed by many authors \cite{5} in studying
the UV behavior of the solution. In a region where
$p^{2}<m^{2}$, but still $m^{2}\gg \Lambda_{\mbox{\scriptsize QCD}}^{2}$
\begin{eqnarray} \label{26}
\bar g^{2}(t)=\bar g^{2}(m^{2})\theta (m^{2}-k^{2}) +
\bar g^{2}(k^{2})\theta (k^{2}-m^{2})
\end{eqnarray}
which combining with (\ref{25}) gives a complete description of
the kernel for the massive quarks in the entire momentum region.

\section{Numerical Solutions}\label{numer}
In above, we have derived a Bethe-Salpeter kernel suitable for heavy
quarks based on the RGE analysis. The same analysis can be made
on the wave function $A(p^{2})$. The leading-logarithmic
corrections to $A(p^{2})$ are proportional to $\bar g^{2}(s)$
where $s=\frac{1}{2}\ln \max (p^{2},m^{2})/\Lambda_{\mbox{\scriptsize
QCD}}^{2}$. However,
when we work in the Landau gauge, these corrections are absent
and the problem can be further simplified. In the Landau gauge,
we have
\begin{eqnarray} \label{27}
A(p^{2})=1\; ,\quad\quad\quad Z_{\psi}=1
\end{eqnarray}
and the anomalous dimension $\gamma _{F}$ is equal to zero.
Substituting the kernel and (\ref{27}) into the SD equation, we
obtain two coupled integral equations which are valid in
different regions respectively. We would like to observe how
the effective gauge interaction affects the quark self-mass
when the quark current mass changes. Thus it is instructive to
measure all dimensionful quantities in unit of the on-shell
current mass $m$, i.\ e.\ we define new variables
\begin{eqnarray} \label{28'}
x=\frac{p^{2}}{m^{2}}\; ,\quad y=\frac{k^{2}}{m^{2}}\; ,
\quad \Delta^{2}=\frac{\Lambda^{2}}{m^{2}}\; , \quad
B(x)=\frac{B\left( \frac{p^{2}}{m^{2}}\right)}{m}\; .
\end{eqnarray}
Clearly, if there are no interactions, $B(x)=1$. In addition,
we are interested in the self-mass function at the scale of the
on-shell current mass. Thus we choose the renormalization point
$\mu =m(\mu=m)=m$. Putting everything together, we obtain the
following integral equations in the Euclidean space
\begin{eqnarray} \label{28}
B(x) & =& B_{0}(\Delta )+\frac{\lambda_{0}}{x}\int
_{0}^{x}\mbox{d}y\frac{yB(y)}{y+B^{2}(y)} \\
& & +\lambda_{0}\int
_{x}^{1}\mbox{d}y\frac{B(y)}{y+B^{2}(y)} + \int
_{1}^{\Delta^{2}}\mbox{d}y\lambda (y)\frac{B(y)}{y+B^{2}(y)}
\quad\quad\quad\quad \mbox{($x<1$)}\nonumber
\end{eqnarray}
\begin{eqnarray} \label{29}
B(x) & =& B_{0}(\Delta )+\frac{\lambda (x)}{x}\int
_{0}^{1}\mbox{d}y\frac{yB(y)}{y+B^{2}(y)} \\
& & +\frac{\lambda (x)}{x}\int
_{1}^{x}\mbox{d}y\frac{yB(y)}{y+B^{2}(y)} + \int
_{x}^{\Delta^{2}}\mbox{d}y\lambda (y)\frac{B(y)}{y+B^{2}(y)}
\quad\quad\quad\quad \mbox{($x>1$)}\nonumber
\end{eqnarray}
where
\begin{eqnarray} \label{30}
\lambda_{0}=\frac{d}{\ln m^{2}/ \Lambda_{\mbox{\scriptsize QCD}}^{2}}\; &,&
\quad\quad \lambda
(x)=\frac{d}{\ln
x+\ln m^{2}/\Lambda_{\mbox{\scriptsize
QCD}}^{2}}=\frac{d\lambda_{0}}{\lambda_{0}\ln x+d}\\
B_{0}(\Delta )&=&\left(
\frac{\ln{m^{2}} /{\Lambda_{\mbox{\scriptsize QCD}}^{2}}}{\ln
{\Lambda^{2}}/{\Lambda_{\mbox{\scriptsize QCD}}^{2}}}\right)
^{d}=\left( \frac{d}{\lambda_{0}\ln \Delta^{2}+d}\right)^{d}\;
\nonumber
\end{eqnarray}
where $d=3C_{2}(N)/\beta_{0}$. Eqs.\ (\ref{28}) and (\ref{29})
are coupled equations since
(\ref{28}) contains an integration from $1$ to $\Delta^{2}$
which requires the solution for $x>1$ while (\ref{29}) contains
an integration from $0$ to $1$ which requires the solution for
$x<1$. It is easily seen from (\ref{28}) and (\ref{29}) that
$B(x)$ is continuous at $x=1$ because both equations give the
same limit $B(1)$. The derivative $B^{\prime}(x)$, however, is
not continuous. The discrepancy between the limits from $x<1$
and $x>1$
\begin{eqnarray} \label{b}
\left.\Delta B^{\prime}(x)\right| _{x=1}\equiv
\left.B^{\prime}(x)\right| _{x\rightarrow 1^{-}} - \left.B^{\prime}(x)\right|
_{x\rightarrow 1^{+}}\propto \lambda_{0}^{2}
\end{eqnarray}
is of higher order and the artifact of the leading-logarithmic
approximation that we have used to model the kernel.

When $x\rightarrow\infty$, an asymptotic solution to (\ref{29})
can be obtained analytically as is done by many authors by
converting (\ref{29}) into a differential equation. The
regularity of $B(x)$ when $x\rightarrow\infty$ allows one to
linearize the differential equation in the UV region. The
solution reads
\begin{eqnarray} \label{31}
B(x)=\left(\frac{d}{\lambda_{0}}\right)^{d}\left(
\frac{d}{\lambda_{0}}+\ln x\right)
^{-d}+c_{2}\frac{1}{x}\left(\frac{d}{\lambda_{0}}
+\ln x\right) ^{d-1}\quad\quad\quad
\mbox{($x\rightarrow\infty$)}
\end{eqnarray}
where $c_{2}$ is the coefficient of the regular solution which
cannot be determined by the UV boundary condition. It is
expected that when $x$ becomes small, the regular solution
is important in the self-mass. The first term in the
r.\ h.\ s.\ of (\ref{31}) goes to zero much more slowly than
the regular solution when $x\rightarrow\infty$ and is referred
to as the irregular solution. It completely dominates the
self-mass in the UV
region.

To solve the coupled integral equations (\ref{28}) and
(\ref{29}) numerically in the entire region, clearly, we have to
specify the integration cutoff $\Delta^{2}$ since practically
we cannot reach  $\Delta^{2}=\infty$. This will bring us some
ambiguities of choosing $\Delta$ and we have to estimate the
computation errors with it. Fortunately, the asymptotic
solution (\ref{31}) provides a useful hint that how the
solution should behave when $x$ is very large. We can actually
use the irregular solution to evaluate the integration from a
very large $x$, say $X=\mbox{e}^{40}$, to $\Delta^{2}$ both in
(\ref{28}) and in (\ref{29})
\begin{eqnarray} \label{32}
\int^{\Delta^{2}}_{\mbox{e}^{40}}\mbox{d}y \lambda
(y)\frac{B(y)}{y+B^{2}(y)}=-\left(\frac{d}{\lambda_{0}}\right)^{d}\left[\left(
\frac{d}{\lambda_{0}}+\ln \Delta^{2}\right) ^{-d}-\left(
\frac{d}{\lambda_{0}}+40\right) ^{-d}\right]
\end{eqnarray}
and add this term on the inhomogeneous term $B_{0}(\Delta )$.
The $\Delta^{2}$-dependence in the new inhomogeneous term
cancels as can be seen from (\ref{30}) and (\ref{32}). The
accuracy of the computation is controlled by $X$ and the total
number of integration points, and we find $X=\mbox{e}^{40}$ is sufficient
to require the relative errors less than $10^{-3}$. We
further change the momentum variable from $x$ to $\ln (1+x)$ in
order to deal with the huge integration interval from $0$ to
$\mbox{e}^{40}$. The IR behavior of the solution, which we are
particularly interested in, is not
deformed very much from this transformation since $\ln (1+x)$
behaves like $x$ when $x$ is very small. We then integrate a
trial function from $0$ to $40$ and iterate the results till
the desired accuracy is reached.

Figure \ref{fig3} illustrates the numerical solutions to the SD
equation as a function of the momentum and $\lambda_{0}$ or the
current mass $m$. We truncate the graphs at $\ln (1+x)=6$ in
order to make the IR behavior of the solution more obvious. The
discontinuity of the slope at $\ln (1+x)=\ln 2$ where $x=1$ is
present as expected because (\ref{28}) and (\ref{29}) are not
smoothly connected. The corner grows sharper when $\lambda_{0}$
becomes larger which is also predicted by (\ref{b}). The
overall feature of the curves is that the quark self-mass
monotonically decreases as the momentum gets large and
eventually goes to zero when $x\rightarrow \infty$. The speed
is controlled by the current mass, the smaller $m$ is, the
faster it goes down. There is a value of the momentum around
the current mass, above which the self-mass is less than the
current mass while below which the self-mass is enhanced from
the current mass by interactions. The constituent mass defined
as the self-mass at $x=0$ can substantially differ from the
current mass when $m$ is around a few hundred MeV. The
enhancement can be as high as 60\% when $m=1.3\l \simeq 270
\mbox{MeV}$ for example.

There is another point in Figure \ref{fig3} that we would like
to address. The different curves seem to cross at the same
point where $\ln (1+x)=1$ or $x\simeq 1.7$. This can be a matter of
the normalization condition of the self-mass implied in the SD
equation. It by itself is not surprising in the perturbation
theory where we can normalize the renormalized self-mass as
$B_{R}(p,\mu )|_{\mu =\sqrt{1.7}m}=m$. In our case, we do not need to
impose such a condition since we simply look for a solution to
the SD equation. It is likely that the SD equation has
incorporated a similar condition implicitly.

\nonum\section{Acknowledgement}
We wish to thank Jinsheng Hu for his help on writing the
computing programs and Howard Trottier for a suggestion on
the normalization condition. This work has been supported in
part by an operating grant from the Natural Sciences and
Engineering Research Council of Canada.

\figure{The skeleton diagrams of Eq.\ (\ref{7}) and the
perturbative expansion\label{fig1}}
\figure{The perturbative expansion of the Bethe-Salpeter kernel
$K(p,k)$\label{fig2}}
\figure{The quark self-mass functions for massive quarks
\label{fig3}}
\end{document}